\documentstyle[aps,12pt,epsfig]{revtex}
\title{Role of deformation in the decay of $^{56}$Ni and $^{40}$Ca di-nuclei}

\author{ C. ~Bhattacharya${\thanks{Permanent address : Variable Energy
Cyclotron Centre, 1/AF Bidhan Nagar, Calcutta, India}}$, M. ~Rousseau, C.
~Beck,  V. ~Rauch, R. ~Nouicer,  R.M. ~Freeman, O. ~Stezowski, D. Mahboub} 

\address{IReS, UMR7500, CNRS-IN2P3 et Universit\'e Louis Pasteur, F-67037
Strasbourg, France} 

\author{S. ~Belhabib, A. ~Hachem,  E. ~Martin}

\address{ Universit\'e de Nice-Sophia-Antipolis, Nice, France}

\author{ A. ~Dummer, S.J. ~Sanders}

\address{ University of  Kansas,Lawrence, Kansas 66045,USA}

\author{ A. ~Szanto de Toledo}

\address{Instituto de Fisica da Universidade de S\~ao Paulo, S\~ao Paulo,
Brazil}

\begin{document}
\maketitle

\newpage

\centerline{\bf ABSTRACT}

\vskip 2.0cm

\begin{abstract}

{\it Inclusive as well as exclusive energy spectra of the light charged
particles emitted in the $^{28}$Si(E$_{lab}=112.6$ MeV) +$^{28}$Si,$^{12}$C
reactions have been measured at the Strasbourg VIVITRON facility in a wide
angular range 15$^0$ - 150$^0$, using the ICARE multidetector array. The
observed $\alpha$-particle energy spectra are generally well reproduced by the
statistical model using a spin-dependent level density parameterisation. The
results suggest significant deformation effects at high spin. } 

\end{abstract} 

\newpage

\section{Introduction}

\vskip 1.2cm

In recent years, a number of experimental and theoretical studies have
been made to understand the decay of light di-nuclear systems (A $\leq$ 60)
formed through low-energy (E$_{lab}$ $\leq$ 10 MeV/nucleon), heavy-ion
reactions. In most of the reactions studied, the properties of the observed,
fully energy damped yields have been successfully explained in terms of either
a fusion-fission (FF) mechanism or a heavy-ion resonance behavior \cite{far}.
The strong resonance-like structures observed in elastic and inelastic
excitation functions of $^{24}$Mg+$^{24}$Mg  \cite{zur} and
$^{28}$Si+$^{28}$Si \cite{bet} have indicated the presence of shell stabilized,
highly deformed configurations in the $^{48}$Cr and $^{56}$Ni compound systems,
 respectively.
In a recent experiment using  EUROGAM, the
present collaboration studied the possibility of preferential population of
highly deformed bands in the symmetric fission channel of the $^{56}$Ni
compound nucleus as produced through the $^{28}$Si+$^{28}$Si \cite{nou} reaction
at E$_{lab} = 111.6$ MeV. The present work aims
to investigate the possible occurence of highly deformed configurations of the
$^{56}$Ni and $^{40}$Ca di-nuclei produced in the $^{28}$Si+$^{28}$Si and
$^{28}$Si+$^{12}$C reactions through the study of light charged particle
(LCP) emission. In-plane coincidences of the LCP's with both evaporation
residues (ER) and FF fragments have been measured. The LCP's emitted from FF
fragments may provide informations on the deformation properties of these
 fragments. Moreover,
the in-plane angular correlations data will be used to extract the 
temperatures of the emitters. In this paper we will concentrate on
 the ER results.

\newpage

\section{Experimental details}

\vskip 1.2cm

The experiments were performed at the IReS Strasbourg VIVITRON tandem
facility using 112.6 MeV $^{28}$Si beams on $^{28}$Si (180 $\mu $g/cm$^2$) and
$^{12}$C (160 $\mu $g/cm$^2$) targets. Both the heavy ions and their associated
LCP's were detected using the {\bf ICARE} charged particle multidetector
array \cite{bel}. The heavy fragments (ER, quasi-elastic, deep-inelastic and FF
fragments) were detected in eight telescopes, each consisting of an
ionization chamber (IC) followed by a 500 $\mu$m Si detector. The in-plane
detection of coincident LCP's was done using four triple telescopes (Si
40 $\mu$m, Si 300 $\mu$m, 2 cm CsI(Tl)) placed at forward angles, 16 two-element telescopes (Si 40 $\mu$m, 2 cm CsI(Tl)) placed at forward and
backward angles and two telescopes consisting of IC's followed by
 500 $\mu$m Si detectors
placed at the most backward angles. The IC's were filled with isobutane and the
pressures were kept at 30 torr and at 60 torr for detecting heavy fragments and
light fragments, respectively. Typical inclusive and exclusive
(coincidence with all ER's detected at 15$^{o}$) energy spectra of
$\alpha$ particles at 40$^0$ for the $^{28}$Si+$^{28}$Si reaction
 are shown by
solid histograms in Fig. 1(a) and 1(b), respectively. Exclusive
$^{28}$Si+$^{12}$C $\alpha$ spectra measured at 40$^0$ in coincidence with S
and P ER's at 15$^o$ are also displayed in Fig. 2. 

\newpage

\section {Experimental results and statistical-model calculations}

\vskip 1.2cm

The data analysis  was performed using  CACARIZO, the Monte
Carlo version of the statistical-model code CASCADE \cite{vies}. The angular
momenta distributions, needed as the principal input to constrain
 the calculations
were taken from compiled $^{28}$Si+$^{28}$Si \cite{vin1} and
$^{28}$Si+$^{12}$C \cite{har,vin2} complete fusion data. The
other ingredients for the realistic statistical-model calculations such as
 the nuclear level densities and the barrier transmission coefficients,
are usually deduced from the study of the evaporated light particle spectra.
In recent years, it has been observed in many cases
 that the standard statistical model cannot predict the shape of the
  evaporated $\alpha$-particle
energy spectra  satisfactorily \cite{vies,rana,govil}, with the measured
 average energies of the
$\alpha$ particles generally much lower than the corresponding theoretical
predictions. Several attempts have been made  to explain
this anomaly either by changing the emission barrier or by
using a spin-dependent
level density. The change in the emission barriers and consequently the
transmission probabilities affects the lower energy part of the calculated
evaporation spectra. On the other hand, the high-energy part of the spectra
depends critically on the available phase space obtained from the level
densities at high spin as well as the corresponding transmission
coefficients. In hot rotating nuclei formed in heavy-ion reactions,
the level density at higher angular momentum should be spin dependent. The
level density, $\rho(E,J)$, for a given angular momentum $J$ and energy $E$ is
given by the well known Fermi gas expression :
\begin{equation}
\rho(E,J) = {\frac{(2J+1)}{12}}a^{1/2}
           ({\frac{ \hbar^2}{2 {\cal J}_{eff}}}) ^{3/2}
           {\frac{1}{(E-\Delta-E_J)^2} }exp(2[a(E-\Delta-E_J)]^{1/2}),
\label{lev}
\end{equation}
where $a$ is the level density parameter, $\Delta$ is the pairing correction
and  E$_J$ = $\frac{ \hbar^2}{2 {\cal J}_{eff}}$J(J+1) is the rotational
energy, ${\cal J}_{eff}= {\cal J}_0 \times (1+\delta_1J^2+\delta_2J^4)$ is
the effective moment of inertia,  ${\cal J}_0$ is the rigid body moment of
inertia
and $\delta_1$, $\delta_2$ are the deformation parameters \cite{vies,govil}. 



By changing the deformation parameters  one can
simulate the deformation effects on the level densities.
 The CACARIZO calculations have been
performed using two sets of input parameters: one with a standard set and
another with non-zero values for the deformation parameters. 


The solid lines in Fig. 1 show the predictions of CACARIZO using the standard
parameter set with the usual liquid drop model deformation.  It is clear
that the average energies of the measured $\alpha$ spectra are lower than those
predicted by the standard statistical-model calculations. The dashed lines show
the predictions of CACARIZO using $\delta_1$ = 3.2 x 10$^{-4}$ and $\delta_2$ =
2.2 x 10$^{-7}$. The shapes of the inclusive as well as the exclusive $\alpha$
energy spectra are well reproduced after  including the deformation effects. 



In the case of $^{28}$Si+$^{12}$C an interesting result is observed.
In order to explain the inclusive energy spectra of $\alpha$-particles it has
been necessary to use the similar  deformation parameters
 as for $^{28}$Si+$^{28}$Si system.
However, it was not possible to explain the
exclusive energy spectra of $\alpha$ particles obtained in coincidence
with all of
the ER's using this set. Therefore, $\alpha$ energy spectra have been generated in coincidence
with individual S and P ER's as shown in Fig. 2. The shape of the $\alpha$ spectrum (solid histograms) obtained in
coincidence with S is completely different from the spectrum obtained in
coincidence with P.

The dashed lines in Fig. 2 are the predictions of CACARIZO using non-zero
values of $\delta_1$ and $\delta_2$. The shape of the $\alpha$ spectrum
measured in coincidence with P is reasonably well reproduced by the theoretical
curve. However, the model could not predict the shape of the $\alpha$ spectrum
obtained in coincidence with S. This is due to the fact that in this case,
there may be additional contributions to the $\alpha$-particle spectra from the
decay of unbound $^8$Be, produced through a binary decay such as asymmetric FF
and/or an orbiting mechanism with $^{40}$Ca $\rightarrow$ $^{32}$S+$^8$Be. This
confirms the double-humped structure found in the inclusive $^{32}$S velocity
spectra measured by Harmon et al. \cite{har}. The question of the real nature
(FF or orbiting) of this decay process remains open.

\newpage

\section{Summary}

\vskip 1.7cm

Inclusive as well as exclusive energy spectra of $\alpha$-particles have been
measured for the reaction $^{28}$Si+$^{28}$Si and $^{28}$Si+$^{12}$C,
respectively. The observed energy spectra of $\alpha$ particles are not well
reproduced by the standard statistical-model calculations with the usual
liquid drop model deformation. However,  a
satisfactory description of the measured energy spectra has been  achieved by
invoking the changes in the level density and barrier due to the onset of large
deformation effects at high spins. The $\alpha$ spectra obtained in coincidence with 
S for the reaction $^{28}$Si +$^{12}$C  have an additional component which
may  come
from the decay of $^8$Be, which is unbound and produced through the binary
decay of $^{40}$Ca $\rightarrow$ $^{32}$S+$^8$Be. Work is in progress to
analyse the proton energy spectra as well as the angular correlations of both
the proton and $\alpha$ in-plane angular correlations. 


\newpage

\section{Figure Captions}

\vskip 1.7cm

\begin{figure} [h]
\caption{ $\alpha$-particle energy spectra
obtained at 40$^{0}$ for the $^{28}$Si+$^{28}$Si reaction.}
\label{fig2}
\end{figure}

\begin{figure} [h]
\caption{Exclusive $\alpha$-particle energy spectra obtained at 40$^{o}$
for the $^{28}$Si+$^{12}$C reaction. }
\label{fig2}
\end{figure}

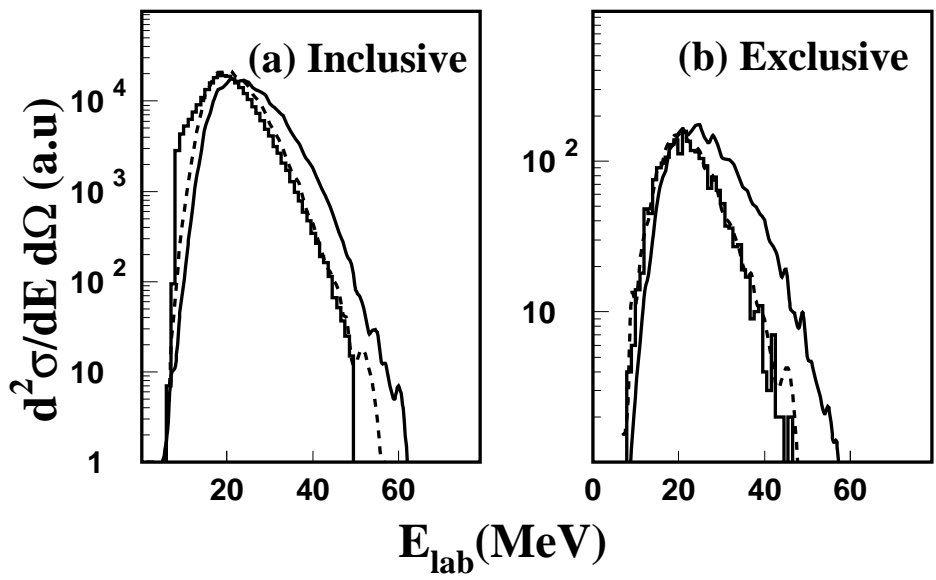
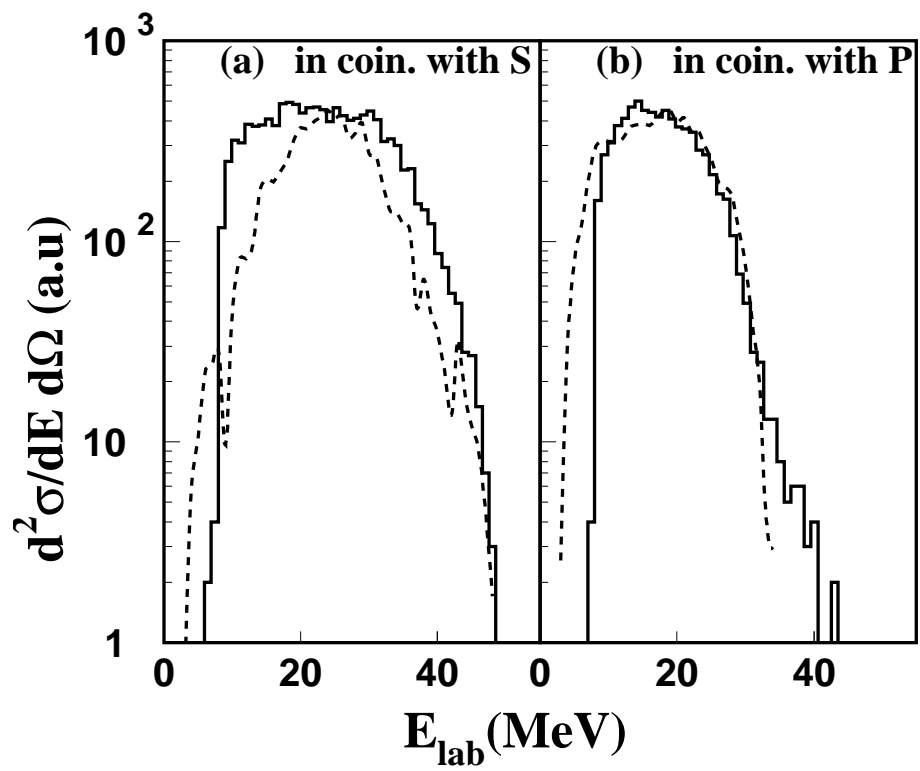
\end{document}